# THE UPGRADE OF J-PARC LINAC LOW-LEVEL RADIO FREQUENCY SYSTEM

S. Li[*], SOKENDAI, Hayama, Kanagawa, 240-0193, Japan
K. Futatsukawa, Y. Fukui, Z. Fang, KEK, Tsukuba, Ibaraki 305-0801, Japan
S. Shinozaki, Y. Sato, S. Mizobata, JAEA, Tokai, Ibaraki, 319-1195, Japan

*Abstract*

The j-parc linac was consist of 324MHz low-$\beta$ section and 972MHz high-$\beta$ section. There is a total of 48 stations. And each station was equipped with an independent LLRF (Low-Level Radio Frequency) system to realize an accelerating field stability of ±1% in amplitude and ±1° in phase [1]. For these llrf system, some of them, especially the 324MHz low-$\beta$ section, had already been used for more than 10 years. Due to lack of supply, it had become more and more difficult to do the system maintain. And in the near future, the beam current of j-parc linac was planned to increase to 60mA. At that time, the current system will face a huge pressure in solving the beam loading effect. Considering these, a new digital llrf system was developing at j-parc linac. In this paper, the architecture of the new system will be reported. The performance of system with a test cavity is summarized.

## INTRODUCTION

J-PARC (Japan Proton Accelerator Complex) LINAC was commissioned in October 2006. Its work is to accelerate an H⁻ beam with a peak current of 50mA and a pulse width of 500us up to 400MeV and then injected to Rapid Cycling Synchrotron (RCS) [2]. Up to present, the low-level radio frequency control system of LINAC has been running for more than 10 years. Even it still works well, the supplier didn't produce the same products many years before, nor did they maintain. Besides this, according to the plan, the beam current of LINAC will increase to 60mA in recent months. As the result, the beam loading effect will be more serious, it puts forward a higher requirement for the performance of LLRF system. Considering these, a new LLRF system for J-PARC LINAC was developed and tested in the last few months, shown as Fig.1.

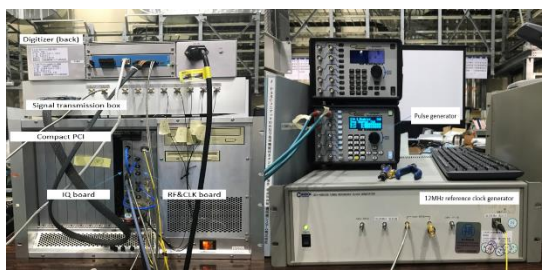

Figure 1: Low-level radio frequency control system

The whole system development will go through several stages. It will take about two years. If we find some shortages on the system, we will correct it in the next version. Now we are just on the first stage. In this stage, the performance testing of digitizer is most important part. The llrf system that shown in the fig.1 we called it "prototype zero". This one is similar with the llrf system that used in STF (Superconducting RF Test Facility). In this upgrade, we still use the compact PCI to hold the RF&CLK board and IQ modulator. In the next stage, we want to integrate all of them into the MTCA.4 chassis. This time the new FPGA was no longer in the compact PCI, but in a digitizer, and it will be introduced in detail in the next part.

[*] The authors of this work grant the arXiv.org and LLRF Workshop's International Organizing Committee a non-exclusive and irrevocable license to distribute the article, and certify that they have the right to grant this license.

## SYSTEM ARCHITECTURE

As shown in the Figure 2, the pulse generator provides 3 trigger signals to the system. Clock generator send 12MHz Square wave and 312MHz optical signal to the RF&CLK board (Mixer). The RF&CLK board on the one hand produce 312MHz and 324MHz intermediate signal then send to IQ board, on the other hand, generate 48MHz clock signal for the digitizer. IQ board can send and receive RF signal to/ from the test cavity. In order to facilitate the communication between the compact PCI and digitizer, a signal transmission chassis was used.

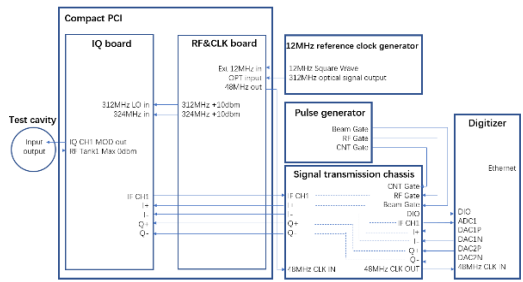

Figure 2: System Test Schematic Diagram

The new digitizer adopted MTCA.4 architecture. It includes 14 ADC, 2 DAC, 2 FPGA and optical communication connector. Both ADC and DAC are 16bits. The sampling speed of them are 80MSPS and 320MSPS (Million Samples per Second). In FMC (FPGA Mezzanine Card) board, Xilinx Spantan6 FPGA was used, which main tasks refer to IQ convert, amplitude and phase adjustment of ADC, vector-sum control and the control & monitoring of high-speed serial signal. Zynq 7000 FPGA was equipped in Carrier board. Its main function is to do the vector-sum control, filtering processing, feedforward & feedback control, PI control, amplitude and phase adjustment of DAC, EPICS input/output control (IOC) and the interlock criterion etc. Figure.3 show the FPGA signal flow in digitizer.

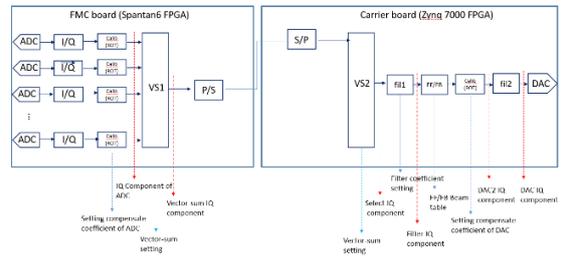

Figure 3: FPGA signal flow diagram

## SYSTEM PERFORMANCE

According to the control theory, the system transfer function for an open-loop system can be written as below,

$$H_{ol}(s) = \left(\frac{K_i K_p + K_p s}{s}\right) \cdot \left(\frac{w_{0.5}}{s+w_{0.5}}\right) \cdot e^{-\tau_d s} \cdot \left(\frac{w_f}{s+w_f}\right)$$

(1)

In our case, the Q factor of test cavity is 3664, half band width of cavity $w_{0.5}$ is $2.78 \times 10^5 rad/s$, system delay $\tau_d$ is 2.38usec, in which 2.11usec (delay from DAC to ADC outside FPGA) + 0.27usec (FPGA delay =13 clock cycles, 1 clock cycle = 1/48MHz). We didn't use filter, so $w_f$=0. In PI control, if we choose $K_p$=1, $K_i$=0, then equation (1) can be written as follow,

$$H_{ol}(s) = \left(\frac{2.78\times 10^5}{s+2.78\times 10^5}\right) \cdot e^{-2.38\times 10^{-6}s} \quad (2)$$

From Bode plot we find that system is stable. And when we choose $K_p$=3, $K_i$=0, we get the critical situation, shown as Figure. 4. It means that $K_p$ should be less than 3 when we do the PI value setting.

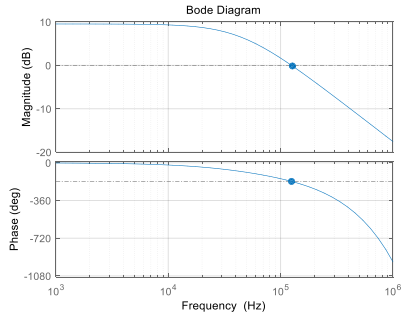

Figure 4: Bode plot of system transfer function when $K_p$=3, $K_i$=0

For our system, P value set to 8 means actual $K_p$ equals to 1. In practice operation, P value should not over 8, or it will cause oscillation. Figure below show the amplitude & phase stabilities of system when P value set to 5 and I value set to 2500 in real system. The amplitude stability is ±0.15%, phase stability is ±0.1°.

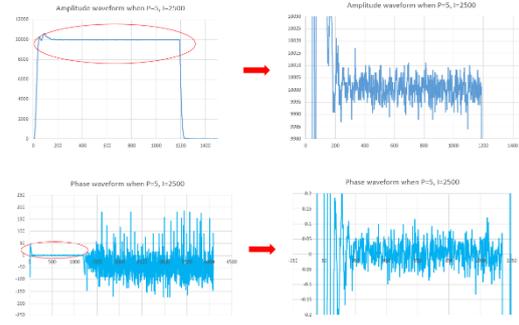

Figure 5: Amplitude & Phase waveform when P value set to 5, I value set to 2500

## SYSTEM DEFECTS AND FUTURE DEVELOPMENT PLAN

In J-PARC, the cycle of chopped beam is 0.815usec, and the feed-forward waveform should be synchronizing with the chopping pulse, shown as Figure. 6. However, the rising-time of DAC that we measured is 3.66usec, shown as Figure. 7. This value is too large to do the feedforward compensation. Additionally, in J-PARC one station just have one or two cavities, we don't need so many ADCs. For the future system, we wish to improve the performance of DAC and ADC, shorten their rising-time and response time. Moreover, current system still use Compact PCI crate and a signal transmission chassis. In the future, we want to integrate all of these function into the MicroTCA.4 chassis.

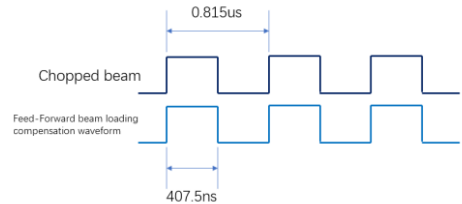

Figure 6: Illustration of chopped beam and feed-forward waveform

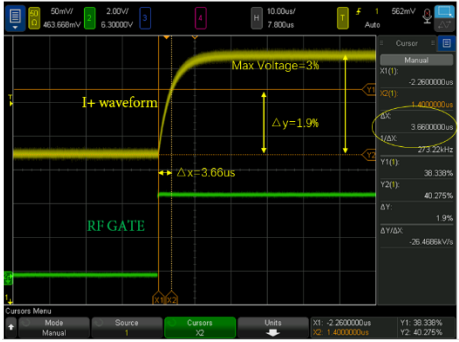

Figure 7: DAC rising-time

## SUMMARY

A new llrf control system was developed in j-parc linac. It adopted MTCA.4 architecture, which allowed FPGA to communicated with EPICS directly. Due to it was just in the first stage of development, a signal transmission box and compact PCI were still used. The system architecture and signal flow in FPGA were introduced. After adjusting the proportional and integral parameter to an ideal value, system can achieve a good stability. By using a test cavity, the experiment result shows that it can realize an RF field stability of ±0.15% in amplitude and ±0.1° in phase. Taking the actual situation into account, the "prototype zero" could not satisfy the requirement of j-parc linac. We will replace the ADC and DAC in digitizer and integrate some part into the MTCA.4 chassis in the future.

## REFERENCE

[1]. Z. Fang et al., "RF feedback control system of the J-PARC Linac", PAC07, 2101-2103, USA.

[2]. K. Hasegawa et al., "Commissioning of the J-PARC Linac", PAC07, 2619-2623, USA.